\documentclass[11pt]{article}
\usepackage{jheppub}
\usepackage{enumitem}
\usepackage{tcolorbox}

\usepackage{jheppub}
\usepackage{mathrsfs}
\usepackage{psfrag}
\usepackage{color}
\usepackage{hyperref}

\usepackage{slashed}
\usepackage{feynmp-auto}
\usepackage{simplewick}
\usepackage{cancel}

\usepackage{todonotes}

\usepackage{amsmath}
\usepackage{amsfonts}
\usepackage{graphicx}
\usepackage{amssymb}
\usepackage{xcolor}
\usepackage{empheq}
\usepackage{mathtools}

\title{Non-vanishing of tidal Love numbers}

\author{Shengsheng Cai$^{*}$}
\author{Kai-Der Wang$^{*}$}

\affiliation{$^*$ Department of Physics and Astronomy, National Taiwan University, Taipei 10617, Taiwan}
\emailAdd{lange2580a@gmail.com}
\emailAdd{kaiderwang@gmail.com}

\abstract{In this note, we analyze black holes solutions under $R^3$ corrections, which is the leading correction induced by quantum corrections in four-dimensions. We showed that perturbations around this black hole background will lead to non-zero tidal Love number (TLN). This further accentuates the ``unnaturalness" of the vanishing TLN for Schwarzschild black hole under Einstein-Hilbert action.}

\begin{document}

\maketitle

\section{Introduction/Motivation}
One of the important observational features of black holes predicted from the Einstein-Hilbert action is the vanishing \emph{tidal Love number} (TLN). The Love number encodes the deformability of stellar objects under external gravitational fields. It was shown in \citep{Binnington:2009bb,Damour:2009vw} that TLN for Schwarzschild solutions~\citep{Schwarzschild:1916uq} vanishes, by studying it for neutron stars and taking the black hole limit. Later on, it was shown by Kol~\citep{Kol:2011vg} that this vanishing result is special to the four-dimensional solution, and is no-longer zero when $D>4$. This brings into light the ``unnaturalness" of TLN in $D=4$, as was emphasized in~ \citep{Porto:2016zng}, where the TNL is mapped to a particular operator in the one-body effective worldline action \citep{Goldberger:2007hy,Rothstein:2014sra},
\begin{equation}\label{t}
S_{eff}=\int d\tau \big (-m+C_{E}E_{\mu\nu}E^{\mu\nu} + C_{B}B_{\mu\nu}B^{\mu\nu}+\cdots
\big )
\end{equation}
The first term is the standard point-particle action, and the second term describe the finite-size effect. Here, $E$ and $B$ are the magnetic and electric component of the Reimann tensor. In principle, we expect the coefficients to be determined by the relevant short-distance scale, in our case $r_{H}$.
As pointed out in \citep{Porto:2016zng} that the vanishing of Love numbers (both electric-type $\emph{k}^{E}$ and magnetic-type $\emph{k}^{B}$) implied that the coefficients $C_{E}$ and $C_{B}$ are zeros at all scales within the EFT realm. Since there is no apparent enhanced symmetry that forces $C_{E(B)} =0$, it from the receiving corrections. Hence, the fact that all of the Love numbers for black holes vanish in classical GR  \textemdash unprotected by symmetries \textemdash  implies  "\emph{fine tunning}" from the EFT point of view \citep{Rothstein:2014sra,Porto:2016pyg}.

Recently, it was shown that under $R^4$ type corrections the TLN is indeed non-zero~ \citep{Cardoso:2018ptl}. There, leading order $R^3$ effects were ignored due to the fact that the presence of such operators in the weak gravity regime, where the cutoff for $R^3$ is suppressed relative to $M_{pl}$, were shown to introduce non-causality~\citep{Camanho:2014apa}. It was argued that such causality is a reflection of the presence of a tower of infinite massive higher spin states at the cutoff scale, with the lightest state inducing long range forces that is in contradiction with observation (see~\citep{Endlich:2017tqa} for details). Note however, that even with just the Einstein-Hilbert action, quantum corrections will introduce log effects associated with  $R^{\mu\nu}_{\hphantom{\mu\nu} \rho\sigma}R^{\rho\sigma}_{\hphantom{\rho\sigma}{\kappa\eta}} R^{\kappa\eta}_{\hphantom{\kappa\eta} {\mu\nu}}$ presented in the Lagrangian which is the two-loop UV divergences in Einstein's gravity \citep{Goroff:1985th}. At low energy these large log effects would dominate, and thus it is still a valid question for how this effects that vanishing of TLN.

Recently, the effects of $R^3$ type operators on classical GR has been initiated, include corrections to gravitational potential~\citep{Brandhuber:2019qpg, Emond:2019crr, Cristofoli:2019ewu}. Here we consider black hole solutions under such deformations, and perturbations around the black hole background. By extrapolating to spatial infinity, we read off the TLN and show that, not surprisingly, it is non-zero. This result further confirms the unnaturalness of vanishing TLN for pure Einstein-Hilbert theory. 

This paper is organized as follows. In section 2 we find a perturbed solution around Schwarzschild black hole under $R^3$ correction. In section 3 we briefly review how to connect coefficients of one-body effective action between TLN and then compute TLN under $R^3$  correction. In section 4 we try to argue the sign of $R^3$ correction. Finally, in section 5, we discuss another possible way to define TLN to make the non-vanishing of TLN more apparent.

\section{The Perturbed Solutions}

Our starting point is the following Lagrangian,
\footnote{Throughout, we work with a positive signature metric and units $c=G=\hbar=1$. Our curvature convention is $R^{\mu}_{\hphantom{\mu}{\nu\rho\sigma}}=\partial_{\rho}\Gamma^{\mu}_{\nu\sigma}+\cdots$ }
\begin{equation}
S = \int d^{4}x \sqrt{-g} \frac{1}{16\pi } \big ( R + \alpha R^{\mu\nu}_{\hphantom{\mu\nu} \rho\sigma}R^{\rho\sigma}_{\hphantom{\rho\sigma}{\kappa\eta}} R^{\kappa\eta}_{\hphantom{\kappa\eta} {\mu\nu}} \big ),
\end{equation}
the field equations derived by extremizing the action $S$ are given by 
\begin{equation}\label{d}
R_{\mu\nu} - \frac{1}{2} R g_{\mu\nu} + \alpha K_{\mu\nu} = 0 ,
\end{equation}
where

\begin{equation}\label{eq:abc}
K_{\mu\nu} = -\frac{1}{2} g_{\mu\nu} (R_{ab}^{\hphantom{ab}{cd}} R_{cd}^{\hphantom{cd}{ef}} R_{ef}^{\hphantom{ef}{ab}}) + 3 (R_{\mu bcd} R^{cdef} R_{ef\nu}^{\hphantom{ef\mu}{b}} -2  \nabla_b \nabla_c R^{c \hphantom{\mu} ef}_{\hphantom{c}{\mu}} R_{ef\nu}^{\hphantom{abc}b})
\end{equation}
In our case the relevant energy scale is only Schwarzschild radius $r_s$, so naive constraint is

\begin{equation}\label{l}
|\alpha|<<r_s^4 
\end{equation}
 thus we can obtain perturbative solutions to the field equations in a power series of $\alpha$. We will take as the starting metric, $\widetilde{g}_{\mu\nu}$, a solution of Einstein equations (Schwarzschild solution) and then, systematically, successive orders of the coupling constant $\alpha$.

To zero order,

\begin{equation}
R_{\mu\nu}(\widetilde{g}_{\mu\nu}) - \frac{1}{2} R(\widetilde{g}_{\mu\nu}) \widetilde{g}_{\mu\nu} = 0
\end{equation}
where

\begin{equation}
\widetilde{g}_{\mu\nu} = - (1 - \frac{2 M}{r}) dt^{2} +(1 - \frac{2 M}{r})^{-1}dr^{2} + r^{2} d\Omega^{2}
\end{equation}

To the first order,

\begin{equation}\label{s}
R_{\mu\nu}(g_{\mu\nu}) - \frac{1}{2} R(g_{\mu\nu}) g_{\mu\nu} + \alpha K_{\mu\nu}(\widetilde{g}_{\mu\nu})= 0
\end{equation}
where to first order in $\alpha$ it is enough to consider in Eq.\eqref{s}  $K_{\mu\nu}(\widetilde{g}_{\mu\nu})$. We find that at first order in $\alpha$ the metric is

\begin{equation}\label{eq:a}
ds^{2} = g_{\mu\nu} dx^{\mu} dx^{\nu} = -f(r) dt^{2} + \frac{1}{g(r)}dr^{2} + r^{2} d\Omega^{2} 
\end{equation}
\begin{equation}
f(r)=1 -\frac{2M}{r}+\alpha\frac{40M^{3}}{r^{7}}
\end{equation}
\begin{equation}
g(r)=1 - \frac{2M}{r} + \alpha( \frac{-392M^{3}}{r^{7}} +\frac{216M^{2}}{r^{6}})
\end{equation}
 Since the spacetime is static, the perturbed horizon is determined by zeros of the metric components $f(r)$ and $g(r)$ defined in Eq. \eqref{eq:a}. By explicit calculation we have verified that $f(r)$ and $g(r)$ share the same zeros at "$\emph{any}$" orders in perturbations. In the first order, the perturbed horizon is located at

\begin{equation}\label{r}
r_{H} = r_{s} + \Delta r =  2 M -\frac{ 5\alpha}{8M^{3}}
\end{equation}
where $r_{s}$ is the Schwarzschild radius.

It is straightforward to calculate the higher order corrections in $\widetilde{g}_{\mu\nu}$, we here only list some results 

\begin{equation}
g_{\mu\nu} = - \Big (1 - \frac{2 M}{r} + \alpha\triangle_{1} + \alpha^{2}\triangle_{2}+ \alpha^{3}\triangle_{3} + \cdots\Big ) dt^{2} + \Big (1 - \frac{2 M}{r} + \alpha \square_{1} + \alpha^{2} \square_{2}+ \alpha^{3} \square_{3} + \cdots\Big )^{-1} dr^{2} + r^{2} d\Omega^{2}
\end{equation}
where

\begin{equation}
\begin{split}
&\triangle_{1} = \frac{40 M^{3}}{r^{7}}  \\
&\triangle_{2} = \frac{127632 M^{5}}{r^{13}} - \frac{1378944M^{4}}{11 r^{12}} + \frac{331776 M^{3}}{11r^{11}} \\
&\triangle_{3} = \frac{532179264 M^{7}}{r^{19}} - \frac{184164516864 M^{6}}{187 r^{18}} + \frac{9263040768 M^{5}}{17 r^{17}} - \frac{93623040 M^{4}}{r^{16}} \\
&\square_{1} = - \frac{392 M^{3}}{r^{7}}+ \frac{216 M^{2}}{r^{6}} \\
&\square_{2} = - \frac{29616 M^{5}}{r^{13}} + \frac{196992 M^{4}}{11 r^{12}} \\
&\square_{3} = \frac{16867008192 M^{7}}{r^{19}} - \frac{4127209873920 M^{6}}{187 r^{18}} + \frac{104591243520 M^{5}}{11 r^{17}} - \frac{1345683456 M^{4}}{r^{16}}
\end{split}
\end{equation}\newline
Here we try to guess the exact solution from the perturbed one. Unfortunately, the $t$ and $r$ components of the metric are not increasing in a regular way. We have no clue what the exact solution is.

\section{BH Tidal Love numbers }
\subsection{Measuring Love}
Here we briefly explain how the coefficients $C_E$ (or $C_B$) in Eq. (\ref{t}) are related to the $k^E_2$ (or $k^B_2$). To illustrate the effective action of tidal interactions, we start with the point-particle worldline action constructed in ref \citep{Chakrabarti:2013lua} reads
\begin{equation}\label{pa}
S_{eff}=\int d\tau \big(   -m-\frac{1}{2}E^{ab}Q_{ab}^{E}-\frac{1}{2}B^{ab}Q_{ab}^{B} +\cdots      \big)
\end{equation}
where $m$ is a parameter of constant mass, $E^{ab}=e^{\phantom{d}a}_\mu e^{\phantom{d}b}_\nu E^{\mu\nu}$, $B^{ab}=e^{\phantom{d}a}_\mu e^{\phantom{d}b}_\nu B^{\mu\nu}$ and $E_{\mu\nu}$($B_{\mu\nu}$) is the electric (magnetic) part of the Weyl tensor $C_{\mu\rho\nu\sigma}$, $E_{\mu\nu}=C_{\mu\alpha\nu\beta} u^{\alpha} u^{\beta}$, $B_{\mu\nu}=\frac{1}{2} \epsilon_{\mu\alpha\beta\rho}   C^{\alpha\beta}_{\phantom{dd}\nu\sigma} u^{\rho} u^{\sigma}$ and $u^\mu$ is the 4-velocity with respect to the proper time $\tau$. The operators $Q_{ab}^{\phantom{dd}E,B}$ are electric and magnetic quadrupole type parity operators composed of worldline degrees of freedom \citep{Chakrabarti:2013xza}. Following the electomagnetic example, the response to an external field can be written as following (after Fourier transform from time to frequency) \citep{Damour:2009vw,Binnington:2009bb,Goldberger:2005cd,Porto:2007qi}
\begin{equation}
\begin{split}
Q_{ij}^{\phantom{ff}E}(w)=-\frac{1}{2}F_{E}(w)E_{ij}(w)\\
Q_{ij}^{\phantom{ff}B}(w)=-\frac{1}{2}F_{B}(w)B_{ij}(w)
\end{split}
\end{equation}
In the low-frequency limit we can expand the function $F(w)$ in powers of $w$,
\begin{equation}
F(w)=2\mu_2+i\lambda w+\mathcal{O}(w^2)
\end{equation}
where the first parameter $\mu_2$ is related to the dimensionless (relativistic, quadrupolar, 2nd-kind) tidal Love number $k_2=\frac{3G\mu_2}{2R^5}$, $R$ is the radius and $G$ is the Newton constant. Furthermore the second term is related to the absorption \citep{Goldberger:2005cd}, and similarly for the magnetic-type components.

In this case, the action in Eq. (\ref{pa}) reduces to \citep{Goldberger:2005cd}
\begin{equation}
S_{eff}=\int d\tau \big(   -m+\frac{\mu_2}{4}E^{ab}E^{ab}+\frac{\mu_2^{\phantom{a}'}}{2}B_{ab}B^{ab} +\cdots      \big)
\end{equation}
which connected with the Eq. (\ref{t}). On the other hand, consider a static, spherically symmetric star of mass $M$ placed in a static external quadrupolar tidal field $\mathcal{E}_{ij}$, the star will develop in response a quadrupole $Q_{ij}$. In the star's ;ocal symptotic rest frame, the asymptotic behavior metric component has the form \citep{Hinderer:2007mb,Thorne:1997kt}
\begin{equation}
g_{tt}=-1+\frac{2M}{r}+\frac{3 Q_{ij}}{r^3}(n^in^j-3\delta^{ij})+\mathcal{O}(\frac{1}{r^3})+\frac{1}{2}\mathcal{E}_{ij}n^in^j r^2+\mathcal{O}(r^2)
\end{equation}
where $n^i=x^i/r$. If we expanded it under spherical harmonic functions, the time-time component reads \citep{Cardoso:2017cfl}
\begin{equation}
g_{tt}=-1+\frac{2M}{r}+\bigg(\frac{2}{r^3}\sqrt{\frac{4\pi}{5}}M_2Y^{20}  +\mathcal{O}(\frac{1}{r^3})        \bigg)+\bigg(-r^2\mathcal{E}_2Y^{20}+\mathcal{O}(r^2)               \bigg)
\end{equation}
where $\mathcal{E}_{ij}=\Sigma_{m=-2}^2\mathcal{E}_{m} Y_{ij}^{2m}$ and $Q_{ij}=\Sigma^2_{m=-2}Q_mY^{2m}_{ij}$, the symmetric traceless tensor $Y_{ij}^{2m}$ are defined by $Y^{2m}(\theta,\phi)=Y_{ij}^{2m}n^i n^j$. And it's similarly for the magnetic-type components
\begin{equation}
g_{t\phi}=\frac{2J}{r}sin^2\theta+\bigg( \frac{1}{r^2}\sqrt{\frac{4\pi}{5}} S_2S_{\phi}^{\phantom{d}20}+\mathcal{O}(\frac{1}{r^2})           \bigg)+\bigg(              \frac{r^3}{3}\mathcal{B}_{2}S_{\phi}^{\phantom{d}20} +\mathcal{O}(r^3)\bigg)
\end{equation}
Hence, we obtain the expression of tidal Love numbers as
\begin{equation}
\begin{split}
k_2^E=-\frac{1}{M^5}\sqrt{\frac{4\pi}{5}}\frac{M_2}{\mathcal{E}_2}\\
k_2^B=-\frac{1}{M^5}\sqrt{\frac{4\pi}{5}}\frac{S_2}{\mathcal{B}_2}
\end{split}
\end{equation}

\subsection{Computing Love}
 In this section, we will analyze the linear perturbation to the BH background solutions to test how robust BH Love number is. The effects of the higher dimension operator come in at two places, first is the deformation of the e.o.m, the other is the BH background.

We linearly perturb around the BH background as,
\begin{equation}
g_{\mu\nu}(\alpha)=g^{BG}_{\mu\nu}(\alpha)+h_{\mu\nu}(\alpha)
\end{equation}
where $g^{BG}_{\mu\nu}(\alpha)$ is the perturbed background spacetime metric defined in Eq.(\ref{eq:a}). We decompose $h_{\mu\nu}(\alpha)$ in spherical harmonics and separate the perturbation into even and odd parts. Again, every function is understood to be dependent on $\alpha$ up to linear order. In the Regge-Wheeler gauge, $h_{\mu\nu}$ can be decomposed as (see Appendix A of \citep{Cardoso:2017cfl} )

\begin{equation}\label{v}
h^{even}_{\mu\nu}=
\begin{bmatrix}
f(r)\widetilde{H}^{lm}_{0}(r)Y^{lm}  &\widetilde{H}^{lm}_{1}(r)Y^{lm}&0&0\\
\widetilde{H}^{lm}_{1}(r)Y^{lm}&\frac{1}{g(r)}\widetilde{H}^{lm}_{2}(r)Y^{lm}
&0&0\\0&0&r^{2}\widetilde{K}^{lm}(r)Y^{lm}&0\\0&0&0&r^{2}sin^{2}\theta \widetilde{K}^{lm}(r)Y^{lm}
\end{bmatrix}
\end{equation}
\begin{equation}\label{o}
h^{odd}_{\mu\nu}=
\begin{bmatrix}
 \phantom{f(r)r}0 & \phantom{f(r)rrrrr} 0&  \phantom{f(r)}\widetilde{h}^{lm}_{0}(r)S^{lm}_{\theta}        &  \phantom{f(r)}\widetilde{h}^{lm}_{0}(r)S^{lm}_{\phi} \\
  \phantom{f(r)r} 0 & \phantom{f(r)rrrrr}0& \phantom{f(r)r}\widetilde{h}^{lm}_{1}(r)S^{lm}_{\theta}& \phantom{f(r)}\widetilde{h}^{lm}_{1}(r)S^{lm}_{\phi}\\
 \phantom{f(r)}\widetilde{h}^{lm}_{0}(r)S^{lm}_{\theta}      & \phantom{f(r)}\widetilde{h}^{lm}_{1}(r)S^{lm}_{\theta}      &0&0\\
\phantom{f(r)}\widetilde{h}^{lm}_{0}(r)S^{lm}_{\phi}& \phantom{f(r)}\widetilde{h}^{lm}_{1}(r)S^{lm}_{\phi}&0& 0
\end{bmatrix}
\end{equation}
where $(S_{\theta},S_{\phi})\equiv (\frac{-Y^{lm},_\phi}{sin\theta},sin\theta Y^{lm},_\theta)$. Since the undeformed Einstein equation would imply that $\widetilde{H}_{0}=\widetilde{H}_{2}\equiv H$ and $\widetilde{H}_{1}=\widetilde{h}_{1}=0$, so we decompose the solutions for the new equation of motion as following :
\begin{equation}
\begin{split}
\widetilde{H}_{0}(r)=H(r)+\alpha  H_{0}(r)   \phantom{ff}\\
\widetilde{H}_{2}(r)=H(r)+\alpha H_{2}(r) \phantom{fe}\\
\widetilde{H}_{1}(r)=\alpha H_{1}(r)\phantom{fjjermej}\\
\widetilde{K}(r)=K(r)+\alpha K_{1}(r)\phantom{ff}\\
 \widetilde{h}_{0}(r)=h(r)+\alpha h_{0}(r)\phantom{eff}\\
 \widetilde{h}_{1}(r)=\alpha h(r)\phantom{fjjrcdmeee}
\end{split}
\end{equation}
Plugging Eq.(\ref{v}) and Eq.(\ref{o}) into Eq.(\ref{d}) and keeping up to first order in $\alpha$, we find that $H_{2}$ can be related to $H_0$ and $H$, and we can show that $\widetilde{H}_1=\widetilde{h}_1=0$, whereas $\widetilde{K}$ can be written as a function of $\widetilde{H}_{0}$ and of the background metric.
The relation is as following. 
\begin{equation}
H_2=\frac{576M(r-3M)H+r^{6}H_0-288Mr(r-2M)H'}{r^6} \phantom{f(r)ddffffffffffffgggffd}
\end{equation}
\begin{equation}
\begin{split}
K_1=\frac{1}{r^7 (r-2M)^2 (-2+l+l^2)}\Big[ 16M(434M^4 +4M^3r(-125+18l(l+1))\phantom{f(r)ddgggggggd}\\+ M^2r^2(283-108l(l+1))+54Mr^3(-2+l+l^2)  -9r^4(-2+l+l^2)   )H  \phantom{fffgggggggg} \\
 +  r(r-2M)\bigg(r^5(-4M^2-2Mr(-4+l+l^2) +r^2(-2+l+l^2))H_0  \phantom{ffgggfffgggggff}\\
 -2M(r-2M)\big(4(53M^2+18Mr(-2+l+l^2) -9r^2(-2+l+l^2))H'-r^6H_0' \big)\bigg)  \Big] \phantom{ffgg}
\end{split}
\end{equation}
Thus in summary, the perturbed metric is parameterized by $(H, h, H_0, h_0)$ where $(H, h)$ would be the linear perturbation solution of the original Einstein-Hilbert action. The function $H_0$ are then constrained by the $(t,t)$ component while $h_0$ from the $(t,\phi)$ component. Using the original equations which $H$ and $h$ satisfied, we can replace higher order derivatives into lower derivatives.

For Electric-type :
\begin{equation}
\begin{split}
H''+\frac{2(r-M)}{r(r-2M)}H'-\frac{4M^2-2l(l+1)Mr+l(l+1)r^2}{r^2(r-2M)^2}H
\phantom{feedffffffffffffffffffffffffff}\\
+\frac{\alpha}{r^8(r-2M)^3 } 
\Bigg\{ 4M\Big[ 11008M^4 +4(-4064+215l(l+1))M^3 r-2(-3756+l(l+1)(443+6l(l+1)))M^2r^2 \phantom{frd} \\
     +12(-90+l(l+1)(22+l+l^2))Mr^3-3l(l+1)(6+l+l^2)r^4        \Big]H 
     +(2M-r)r\Bigg(  r^5(4M^2
     -2l(l+1)Mr \phantom{frdddd}\\
     +l(l+1)r^2)H_0-8M\Big[ 784M^3+3(-183+2l(l+1))M^2r-3(-17+3l(l+1))Mr^2+3(6+l+l^2)r^3         \Big]H'\phantom{frddd} \\ 
         +(2M-r)r\Big[ 2r^5(r-M)H_0'+4M(392M^2-6(35+l+l^2)Mr
     +3(6+l+l^2)r^2)H''+r^6(r-2M)H_0''                \Big] \Bigg)\Bigg\}  =0
\end{split}
\end{equation}

For Magnetic-type :
\begin{equation}
\begin{split}
h''+\frac{4M-l(l+1)r}{r^2(r-2M)}h  \phantom{frdddddddddddddddddddddddddddddddddddddddddddd} \\
-\frac{\alpha}{r^8(r-2M)}\Bigg\{  -4M\Big[ -644M^2+144(1+l+l^2)Mr+3(l-3)l(l+1)(l+4)r^2       \Big]  h            \phantom{frddd}\\
+r\Bigg(r^5\Big(-4M+l(l+1)r\Big)h_0    +4M\Big(   90M(r-2M)h'+r(-142M^2-6(-23+l+l^2)Mr  \\
+3(l-3)(l+4)r^2)h''        \Big)    +(2M-r)r^7h_0''                       \Bigg)                      \Bigg\}=0  \phantom{frdddddddddddddddddddddddddddddddd}
\end{split}
\end{equation}
For $l=2$ the general solutions regular at the horizon are
\begin{equation}\label{i}
\widetilde{H}_{0}=-r^2(1-\frac{2M}{r})\mathcal{E}_2+\alpha\Big(\frac{328M^3-72M^2r-56Mr^2}{r^5}        \Big)\mathcal{E}_2
\end{equation}
\begin{equation}\label{p}
\widetilde{h}_{0}=\frac{1}{3}r^3(1-\frac{2M}{r})\mathcal{B}_2+\alpha\Big(  \frac{-100M^2+60Mr}{3r^3}   \Big)\mathcal{B}_2 \phantom{ffffff}
\end{equation}
From the above solutions Eq. (\ref{i}) and Eq. (\ref{p}), we obtain that (using the convention of \citep{Cardoso:2017cfl})

\begin{empheq}[box=\fbox]{align}\label{uu}
k^E_2 &= 28 \epsilon\\
k^B_2 &= -20 \epsilon
\end{empheq}
We have introduced $\epsilon=\frac{\alpha}{M^4}$ (where $M$ stands for the gravitational mass of spacetime) to make the above quantities dimensionless.

\section{Negative entropy for $\alpha < 0$ ?}\label{ll}

It is straightforward to compute the black hole entropy in higher order gravity using the Wald formula \citep{Wald:1993nt},

\begin{equation}\label{ss}
S = -2 \pi \int_{\Sigma} \frac{\delta\mathcal{L}}{\delta R_{\mu\nu\rho\sigma}}\epsilon_{\mu\nu}\epsilon_{\rho\sigma} ,
\end{equation}
where the integration region $\Sigma$ is the horizon and $\epsilon_{\mu\nu}$ is the binormal to the horizon, normalized so that $\epsilon_{\mu\nu}\epsilon^{\mu\nu} = - 2 $ .

For a spherically symmetric spacetime, the integral in Eq.(\ref{ss}) is easy to calculate, yielding

\begin{equation}
S = -2 \pi A \frac{\delta\mathcal{L}}{\delta R_{\mu\nu\rho\sigma}}\epsilon_{\mu\nu}\epsilon_{\rho\sigma} \Bigg|_{g_{\mu\nu},r_{H}} ,
\end{equation}
where all quantities are evaluated for the perturbed metric $g_{\mu\nu}$ and perturbed horizon radius $r_{H}=r_{s}+ \Delta r$ up to first order in $\alpha$. The perturbed horizon area $A_H$ and the binormal is 
\begin{equation}
A_H = 4\pi r_H^2=16\pi M^2-\frac{10\pi}{M^2}\alpha +\mathcal{O}(\alpha^2)
\end{equation}
\begin{equation}
\epsilon_{\mu\nu}(r) = \sqrt{\frac{f(r)}{g(r)}}(\delta^{t}_{\mu}\delta^{r}_{\nu} - \delta^{r}_{\mu}\delta^{t}_{\nu})
\end{equation}
Expanding the area $A_{H}=A_{s}+\Delta A$ and the Lagrange $\mathcal{L}=\mathcal{\widetilde{L}}+\Delta\mathcal{L}$ in perturbations, we obtain

\begin{equation}\label{m}
S=-2\pi \Big (A_{s} \frac{\delta \mathcal{\widetilde{L}}}{\delta R_{\mu\nu\rho\sigma}} + A_{s}\frac{\delta \Delta\mathcal{L}}{\delta R_{\mu\nu\rho\sigma}} + \Delta A \frac{\delta\mathcal{\widetilde{L}}}{\delta R_{\mu\nu\rho\sigma}} + \mathcal{O}(\alpha^{2})             
 \Big ) \epsilon_{\mu\nu}\epsilon_{\rho\sigma} \Bigg |_{g_{\mu\nu},r_{H}}
\end{equation}
where $\mathcal{\widetilde{L}}$ is the Einstein-Hilbert action and $\Delta\mathcal{L}$ is the counterterm. The first term in Eq. (\ref{m}) is readily obtained

\begin{equation}
\frac{\delta \mathcal{R}}{\delta R_{\mu\nu\rho\sigma}}=\frac{1}{16\pi }g^{\mu\rho}g^{\nu\sigma}
\end{equation}
thus

\begin{equation}
S_{1}=\frac{A_{s}}{4}
\end{equation}
which $A_{s}$ is the original Schwarzschild horizon area.

The appearance of the second term in Eq. (\ref{m}) comes from the$R^{\mu\nu}_{\hphantom{\mu\nu} \rho\sigma}R^{\rho\sigma}_{\hphantom{\rho\sigma}{\kappa\eta}} R^{\kappa\eta}_{\hphantom{\kappa\eta} {\mu\nu}}$ correction, which is responsible for canceling the divergence for graviton loop corrections. It is not hard to compute, we obtain

\begin{equation}
\frac{\delta {R^{ab}_{\hphantom{ab} c d}R^{cd}_{\hphantom{cd}{ef}} R^{ef}_{\hphantom{ef} {ab}} }}{\delta R_{\mu\nu\rho\sigma}}= \frac{3\alpha}{16\pi } g^{a\mu}g^{b\nu} R^{\rho\sigma}_{\hphantom{\rho\sigma}{ef}}R^{ef}_{\hphantom{ef}{ab}}
\end{equation}
since we are working in first order in perturbations, it should be evaluated on the unperturbed metric and horizon radius.
Thus we have :

\begin{equation}
S_{2}=\frac{3 A_{s}}{16M^{4}}\alpha
\end{equation}
The last term in Eq. (\ref{m}) is due to the shift of the horizon radius and we have

\begin{equation}
\Delta A= A_{H}-A_{s}=-\frac{10\pi}{M^{2}}\alpha
\end{equation}
therefore we obtain

\begin{equation}
S_{3}=\frac{\Delta A}{4} =-\frac{5\pi}{2M^2}\alpha=-\frac{5A_s}{32M^4}\alpha
\end{equation}
Thus the total entropy up to first order is

\begin{equation}
S_{total}=S_{1}+S_{2}+S_{3}=\frac{A_s}{4}+\frac{A_s}{32M^4}\alpha
\end{equation}

If $\alpha$ is positive, it seems that everything is fine. The entropy of black  hole is increasing due to the $R^{\mu\nu}_{\hphantom{\mu\nu} \rho\sigma}R^{\rho\sigma}_{\hphantom{\rho\sigma}{\kappa\eta}} R^{\kappa\eta}_{\hphantom{\kappa\eta} {\mu\nu}}$ correction. If $\alpha$ is negative, however, we are probably facing a negative entropy of BH. In the following,  
we will consider the situation when the mass of the black hole is small enough to drive the total entropy become negative and try to argue the sign of $\alpha$.

The validity of the perturbed metric \eqref{eq:a} will be assured if the condition Eq.(\ref{l}) holds. As for the negativity of the BH entropy, we have
\begin{equation}\label{tt}
S_{total} <0 \Rightarrow r_s^{4} < 2 |\alpha|
\end{equation}
Unfortunately, after comparing Eq.(\ref{l}) and Eq.(\ref{tt}), we find that the perturbed metric \eqref{eq:a} is out of the valid regime when $S_{total}$ is negative.

Since the coupling constants in front of higher-dimension operators are important low-energy probe of the ultraviolet completion of general relativity, the bound of these coefficients are of particular interest for us. For example, ref. \citep{Cheung:2018cwt} try to argue that  one sign of the entropy correction from higher-curvature terms would violate the "weak gravity conjecture". Here we try to argue the sign of $\alpha$ coefficient. Unfortunately, the Schwarzschild black hole is not thermodynamically stable \citep{Davies:1978mf}, thus the method in ref. \citep{Cheung:2018cwt}  is not applicable to our case. Moreover, ref. \cite{Wall:2015raa} has proved that for all linear perturbations to Killing horizons will obey a 2nd law (even in cases where the entropy is negative, such as GR with G $<$ 0). So to get a violation we need to go beyond linear order. This could be quadratic order in the perturbation (e.g. gravitons falling into the BH), or it could be nonperturbative (e.g. colliding BH's or formation from collapse \citep{Sarkar:2010xp}). Nevertheless, a negative BH entropy still makes it impossible to interpret it as a statistical (von Neumann) entropy of BH microstates ! Instead of going beyond linear order, we consider simpler situation when the mass of BH is small enough to drive total entropy being negative.

However, there is another possible interpretation of the negative entropy for small black holes, which is that any physical theory with an cubic correction must also have other corrections, such that the small black hole are outside the regime of validity of the Einstein plus cubic theory. For example, \citep{Camanho:2014apa} argued that Gauss-Bonnet (GB) and $R^3$ terms that affect the graviton 3-point interaction lead to causality violations unless the $\alpha$ coefficient is very small (i.e. its scale is set by either qunatum or higher-spin/string corrections). Although the GB issue requires $D > 4$, but the $R^3$ term still comes up in $ D = 4$. And in this case one expects an infinite series of corrections that go beyond  R-cube. This issue actually occurs for \textit{either} sign of $\alpha$. 

\section{Conclusion and Outlook}
In this paper, we show that the leading higher derivative corrections to the Einstein-Hilbert action, $R^3$, will induce non-vanishing TLN. We've also considered the effect of such corrections to the Wald entropy, and suggest a possible bound on the sign of $R^3$ that are absent from any S-matrix arguments. The nonvanishing of tidal Love numbers in higher curvature correction (see also ref. \citep{Cardoso:2018ptl}) further enhanced the unnatural of BH Love numbers being zero in classical GR from the point of effective field theory. Recently, ref. \citep{Penna:2018gfx} make a key observation that membrane fluid velocity being zero \citep{PhysRevD.33.915} is the consequence of the emergent Carroll symmetry near horizon. They further connect between Carroll symmetry and TLN. However, our BH solution also retains Carroll symmetry, and hence the connection with Carroll symmetry requires further investigation. 

From the on-shell point of view, the TLN can be read off from the four-point Compton amplitude of massive scalar and two gravitons. Indeed, operators in the one-body effective action that are linear in the Reimann tensor manifest itself as three-point amplitude where two massive (spinning) particles are coupled to a graviton~\citep{Chung:2018kqs,Guevara:2018wpp}. Operators with two Reimann tensors, including $E^2, B^2$ that encodes the TLN, appears as contact term contribution to the Compton amplitudes. From this point of view, the fact that $R^3$ corrections will modify the Compton amplitude due to the change in the residue of the massless pole, naturally leads to a non-vanishing.\footnote{We thank Yu-tin Huang for pointing this out.} It will be extremely interesting to find a completely on-shell definition of TLN from the Compton amplitude to make this precise . 

\paragraph{Acknowledgements} $\phantom{a}$ \newline
We'd like to thank  Clifford Cheung, Chiang-Mei Chen, Yi-Zen Chu, Ming-Zhi Chung, Barak Kol,  Ira Z. Rothstein, Aron Wall for useful discussion. We would alos like to thank especially Yu-tin Huang for very valuable comments. KDW is supported by MoST Grant No. 106-2628-M-002-012-MY3.

\newpage

\bibliographystyle{JHEP}
\bibliography{mybib.bib}{}
\nocite{*}

\end{document}